\documentclass[12pt]{article}
\usepackage{amsmath,amssymb}

\makeatletter

\newcommand{\re}{\mathbb{R}}

\@addtoreset{equation}{section}
\makeatother

\begin{document}

\thispagestyle{empty}

\setcounter{page}{0}

\mbox{}

\begin{center} {\bf \Large  Can (Electric-Magnetic) Duality Be Gauged?}

\vspace{1.6cm}

Claudio Bunster$^{1,3}$ and Marc Henneaux$^{1,2,3}$

\footnotesize
\vspace{.6 cm}

${}^1${\em Centro de Estudios Cient\'{\i}ficos (CECS), Casilla 1469, Valdivia, Chile}

\vspace{.1cm}

${}^2${\em Universit\'e Libre de Bruxelles and International Solvay Institutes, ULB-Campus Plaine CP231, B-1050 Brussels, Belgium} \\

\vspace{.1cm}

${}^3${\em Max-Planck-Institut f\"ur Gravitationsphysik (Albert-Einstein-Institut),
M\"uhlenberg 1, D-14476 Potsdam, Germany}

\vspace {15mm}

\end{center}
\centerline{\bf Abstract}
\vspace{.8cm}
\noindent

There exists a formulation of the Maxwell theory in terms of two vector potentials, one electric and one magnetic.  The action is then manifestly invariant under electric-magnetic duality transformations, which are rotations in the two-dimensional internal space of the two potentials, and local. We ask the question: can duality be gauged? The only known and battled-tested method of accomplishing the gauging is the Noether procedure.  In its decanted form, it amounts to  turn on the coupling by deforming the abelian gauge group of the free theory, out of whose curvatures the action is built,   into a non-abelian group which becomes the gauge group of the resulting theory. In this article, we show that the method cannot be successfully implemented for electric-magnetic duality. We thus conclude that, unless a radically new idea is introduced, electric-magnetic duality cannot be gauged. The implication of this result for supergravity is briefly discussed.
 
\newpage

\section{Introduction}

\setcounter{equation}{0}

The invariance of the vacuum Maxwell equations under an electric-magnetic duality transformation, i.e., an internal rotation in the two-dimensional plane of the electric and magnetic fields, continues to be a fascinating and inspiring subject.   There is a formulation of the Maxwell theory in terms of two vector potentials, one electric and one magnetic.  The action is then manifestly invariant under electric-magnetic duality transformations, which are rotations in the two-dimensional internal space of the two potentials, and local. It is then natural to ask the question: can duality be gauged, i.e., can the angle of the $SO(2)$ duality rotations be made spacetime dependent? In this article, we attempt to apply  the only known and battled-tested method of accomplishing the gauging, namely the Noether procedure, and show that it is not viable.  Therefore, we answer the question in the negative.

The plan of the paper is the following.  In section \ref{ElMadDu}, we begin by recalling the formulation of the Maxwell theory in terms of two abelian vector potentials ${\mathbf A}_a$, $a=1,2$.  The curl of ${\mathbf A}_1$ is the ordinary magnetic field while that of ${\mathbf A}_2$ is the negative of the electric field.  We emphasize that $SO(2)$ duality rotations in the $a=1,2$ plane are a symmetry of the action and not just of the equations of motion, as it is often incorrectly stated. The action is also invariant under the group $[U(1)]^2$ of independent gauge transformations of each of the two potentials. Next, in that same section, we extend the theory to $n$ copies of the Maxwell field and recall that the global electric-magnetic duality group, which was $SO(2)$ when $n=1$, becomes $U(n)$. The gauge group becomes $[U(1)]^{2n}$. We end section \ref{ElMadDu} by analyzing the subgroups of $U(n)$ which are relevant for the gauging procedure which is discussed next.

Section  \ref{NoYang-Mills} is devoted to applying  the Noether gauging procedure to electric-magnetic duality.  In its decanted form, the procedure amounts to  turn on the coupling by deforming the $[U(1)]^{2n}$ gauge group of the free theory, out of whose curvatures the action is built,   into a non-abelian group which becomes the gauge group of the resulting theory.  The $2n$ generators of the adjoint representation of this gauge group should: (i) be linear combinations of the $n^2$ generators of $U(n)$ in its $(2n)$- dimensional representation; and (ii) in addition, the linear combination should contain the generators of the electric-magnetic duality subgroup $[SO(2)]^n$.    We show that already the first requirement cannot be met and conclude that, unless a radically new idea is introduced, electric-magnetic duality cannot be gauged. This conclusion remains valid if one admits into the action Maxwell terms of the traditional one-potential type. Additionally, in that same section, we discuss some group theory relationships which underlie the previous analysis and end by explaining how the lack of a doubled potential reformulation of the pure nonabelian Yang-Mills action, which is local and manifestly duality invariant,  relates to our results.

In section \ref{Supergravity} we make some brief comments on the implications of our results for supergravity.  Section \ref{Conclusions} is devoted to concluding remarks. Finally we point out, in a Note added, that the analysis and conclusions remain valid when the Maxwell field is coupled to scalar fields in the extended manifestly duality invariant formulation given in \cite{Bunster:2011aw}.

\section{Electric-Magnetic Duality}

\label{ElMadDu}
\setcounter{equation}{0}
\subsection{$SO(2)$-Duality}
The electric-magnetic duality transformation is an internal rotation in the two-dimensional plane of the electric and magnetic fields, 
\begin{eqnarray}
{\mathbf E} \; \; &\rightarrow& \; \;  \cos \alpha \, {\mathbf E} - \sin \alpha \,  {\mathbf B} \label{deltaE0}\\
{\mathbf B} \; \; &\rightarrow& \; \;  \sin \alpha \,  {\mathbf E} + \cos \alpha \, {\mathbf B}. \label{deltaB0}
\end{eqnarray}
In covariant form it reads,
 \begin{eqnarray}
F^{\mu \nu} \; \; &\rightarrow& \; \;  \cos \alpha \, F^{\mu \nu} - \sin \alpha \;  ^*\!F^{\mu \nu} \\
 ^*\!F^{\mu \nu} \; \; &\rightarrow& \; \;  \sin \alpha \,  F^{\mu \nu} + \cos \alpha \; ^*\!F^{\mu \nu}.
\end{eqnarray}

It is often incorrectly stated that duality is only a symmetry of the equations of motion, but this is not the case.  Indeed, as shown in \cite{Deser:1976iy}, duality leaves invariant the standard  Maxwell action. The duality invariance of the action becomes manifest if one introduces a second vector potential by solving the Gauss law for the electric field.  The manifestly duality invariant action reads \cite{Deser:1976iy,Deser:1997mz,Deser:1997se}
\begin{equation}
S^{\hbox{{\tiny inv}}}[A^a_i] = \frac{1}{2}  \int dx^0 \, d^3x \left( \epsilon_{ab} {\mathbf B}^a \cdot \dot{{\mathbf A}}^b  - \delta_{ab} {\mathbf B}^a \cdot {\mathbf B}^b \right), \; \; a, b = 1,2 \label{Action0}
\end{equation}
where 
$$ {\mathbf B}^a  = {\mathbf \nabla} \times {\mathbf A}^a$$ and $\epsilon_{ab} = - \epsilon_{ba}$ is the Levi-Civita tensor in 2 dimensions (with $\epsilon_{12}  = 1$).
The duality rotations in this formulation are
\begin{eqnarray}
{\mathbf A}^1 \; \; &\rightarrow& \; \;  \cos \alpha \, {\mathbf A}^1 - \sin \alpha \,  {\mathbf A}^2 \\
{\mathbf A}^2 \; \; &\rightarrow& \; \;  \sin \alpha \,  {\mathbf A}^1 + \cos \alpha \, {\mathbf A}^2.
\end{eqnarray}
The action (\ref{Action0}) is clearly invariant under these transformations since both $\epsilon_{ab}$ and $\delta_{ab}$ are invariant tensors for $SO(2)$. The standard magnetic field $ {\mathbf B}$  is $  {\mathbf B} \equiv {\mathbf B}^1$ while the standard electric field $ {\mathbf E}$ is  $ {\mathbf E} \equiv - {\mathbf B}^2$.  The vector $ {\mathbf A}^1$ is the standard vector potential $ {\mathbf A}$. The two magnetic fields are canonically conjugate, in the sense that their equal time Poisson brackets are
$$ [B_a^i(x), B_b^j(x')] = \delta_{ab} \epsilon^{ijk}\delta_{,k}(x,x'). $$

\subsection{$U(n)$-Duality}
\label{2.2}
{}For  $n$  Maxwell fields, the manifestly duality invariant action takes the form (\ref{Action0}) but the internal indices run now from $1$ to $2n$ values and the $\epsilon$-symbol is replaced by the antisymmetric canonical symplectic form $\sigma_{MN}$ ($M, N = 1,\cdots, 2n$),
$$ \sigma = \left(\begin{array}{ccccccc} 0 & 1 & 0 & 0 & \cdots & 0 &0 \\ -1 & 0 & 0 & 0 & \cdots & 0&0\\ 0 & 0 & 0 & 1 & \cdots & 0&0\\ 0 & 0 & -1& 0& \cdots & 0&0 \\ \vdots&\vdots&\vdots& \vdots&\vdots & \vdots & \vdots\\ 0&0&0&0 &\cdots & 0 & 1 \\ 0&0&0&0& \cdots &-1 & 0\end{array} \right), $$
which gives
\begin{equation}
S^{\hbox{{\tiny inv}}}[A^M_i] = \frac{1}{2}  \int dx^0 \, d^3x \left( \sigma_{MN} {\mathbf B}^M \cdot \dot{{\mathbf A}}^N  - \delta_{MN} {\mathbf B}^M \cdot {\mathbf B}^N \right). \label{Action1}
\end{equation}

{}For later purposes, it is convenient to introduce the time component $A_0^M$ of the vector potentials by replacing $\dot{A}^M_i$ by $$f_{0i}^M = \dot{A}^M_i - \partial_i A^M_0$$ and to rewrite the action (\ref{Action1}) as
\begin{equation}
S^{\hbox{{\tiny inv}}}[A^M_\mu] = \frac{1}{4}  \int dx^0 \, d^3x \left( \sigma_{MN}\epsilon^{ijk} f_{ij}^M \, f_{0k}^N  - \delta_{MN} f_{ij}^M \, f^{N\, ij} \right).
\label{Action2}
\end{equation}
Because of the Bianchi identity, the term $\partial_k A^M_0$ involving $A_0^M$ in (\ref{Action2}) is a total derivative, so that (\ref{Action1}) and (\ref{Action2}) differ by a surface term and are equivalent.  The time components of the vector potentials occur only through a total derivative and yield the trivial equations $0=0$.  The virtue of writing the action in this form is that its gauge invariances,
\begin{equation}
A^M_\mu \rightarrow A^M_\mu + \partial_\mu \Phi^M, \label{abeliangauge}
\end{equation}
are manifest as the action is expressed in terms of the abelian curvatures $f_{\mu \nu}^M$ only. The gauge symmetry group corresponding to (\ref{abeliangauge}) is  $[U(1)]^{2n}$, one $U(1)$ for each vector potential. There is, however, an invariance larger than (\ref{abeliangauge}): since $A_0^M$ appears in the action only through a total derivative, it can be shifted arbitrarily and not just by the amount $\partial_0 \Phi^M$ but this will only be used tangentially in subsection \ref{Noether} below. 

In addition to local gauge symmetries,  the action possesses also rigid duality symmetries. The duality group clearly contains $n$ factors $SO(2)\times SO(2) \times \cdots \times SO(2)$, namely, one $SO(2)$ for each pair $(A^{2k-1},A^{2k})$ describing a single standard Maxwell field. But duality is in fact bigger, because one can also perform linear transformations that mix the vector potentials belonging to different pairs. Hence  the duality group is enlarged from $[SO(2)]^n$ to $U(n)$ \cite{Gaillard:1981rj,deWit:2001pz}.   This can be seen as follows: linear transformations of the potentials $A^M$,
\begin{equation} 
A^M \rightarrow A'^M = \Lambda^M_{\; \; N} A^N
\end{equation}
where $\Lambda \in GL(2n,\re)$, leave the action (\ref{Action1}) invariant if and only if they preserve the symplectic product and the scalar product (in order to preserve the kinetic term and the Hamiltonian, respectively),
\begin{equation} 
\Lambda^T \sigma \Lambda = \sigma, \; \; \; \Lambda^T I \Lambda = I.
\end{equation}
The first condition implies $\Lambda \in Sp(2n, \re)$, while the second implies $\Lambda \in O(2n)$ ($O(2n)$ always means here $O(2n, \re)$).  Accordingly, the transformation $\Lambda$ must belong to $Sp(2n,\re) \cap O(2n)$, which is the maximal compact subgroup of the symplectic group $Sp(2n,\re)$, known to be isomorphic to $U(n)$ (see, for example, \cite{Helgason}).

In infinitesimal form, the invariance condition reads, with $\Lambda = I + \lambda$,
\begin{equation} 
\lambda^T \sigma + \sigma \lambda = 0, \; \; \; \lambda^T + \lambda = 0,
\end{equation}
or in component form,
\begin{equation} 
 \sigma_{PN} \lambda^P_{\; \; M}  +\sigma_{MP} \lambda^P_{\; \; N} = 0, \; \; \; \delta_{PN} \lambda^P_{\; \; M}  +\delta_{MP} \lambda^P_{\; \; N} = 0. \label{InvarianceCond}
\end{equation}

To study some of the interesting subgroups of $U(n)$, it is convenient to adopt a notation in which the vector potentials $A^{a}_\mu$ are grouped in electric-magnetic pairs, $A^{M}_\mu$, $M = 1, \cdots, 2n$  $\rightarrow A^{Ia}_\mu$,  $I = 1, \cdots n$, $a = 1,2$.   In this notation, the action reads
\begin{eqnarray}
S^{\hbox{{\tiny inv}}}[A^{Ia}_\mu] &=& \frac{1}{4}  \int dx^0 \, d^3x \left( \epsilon_{ab}\delta_{IJ} \epsilon^{ijk} f_{ij}^{Ia} \, f_{0k}^{Jb}  - \delta_{ab} \delta_{IJ} f_{ij}^{Ia} \, f^{Jb\, ij} \right), \nonumber \\
&& \hspace{2cm} I,J = 1, \cdots n, \; \; \;  a, b = 1,2
\nonumber
\end{eqnarray}

The subgroup $[SO(2)]^n$ of independent $SO(2)$ rotations in the $n$ two-dimensional planes spanned  by $A^{I1}$ and $A^{I2}$ ($I = 1, \cdots, n$)
$$  A'^{Ia} = \!^{(I)}R^a_{\; \; b} A^{Ib},$$ with  $\!^{(I)}R^a_{\; \; b} \in SO(2)$ for each $I$, is the group of strict electric-magnetic duality rotations.  Its diagonal subgroup $SO(2)$ in which one makes the same rotation for each $I$  
$$  A'^{Ia} = R^a_{\; \; b} A^{Ib}, \; \; \;  R^a_{\; \; b} \in SO(2)$$ is the center of $U(n)$.

Another subgroup of $U(n)$ is  the group $SO(n)$ of transformations acting in the same way on $A^{I1}_\mu$ and $A^{I2}_\mu$, i.e.,
$$ A'^{Ia} = R^I_{\; \; J} A^{Ja}, \; \; \;  R^I_{\; \; J} \in SO(n).$$
Only this subgroup survives in the second order formalism where each Maxwell field is described by a single vector potential.  Any attempt to exhibit the full $U(n)$ would require the introduction of non-local kernels.   So, while the standard second order formulation of a collection of Maxwell fields makes manifest only an $SO(n)$ global invariance group, the formulation based on the duality invariant first order action exhibits a $U(n)$ duality symmetry.

\section{Gauging Duality}

\label{NoYang-Mills}
\setcounter{equation}{0}

\subsection{Statement of the problem}

The dimension of the global duality group $U(n)$ is $n^2$, while the number of vector fields is $2n$ which is smaller than $n^2$ for $n > 2$, with equality for $n=2$.  This paper is devoted to the question of whether one can gauge a subgroup of $U(n)$, i.e., turn some of the global duality transformations into local gauge transformations.  To achieve this goal: 
(i) One can either take advantage of the vector fields that are already present; or  (ii) One may add new vector fields which may be either described by the same duality-invariant action,  thereby simply  increasing the range of the index $M$, or by the standard Maxwell action $-\frac{1}{4} \int d^4 x f_{\mu \nu} f^{\mu \nu}$. 

Thus, we take as starting point the general action
\begin{eqnarray}
&&S[A^\Delta_\mu] = S^{\hbox{{\tiny inv}}}[A^M_\mu] + S^{\hbox{{\tiny Max}}}[A^{(\alpha)}_\mu], \label{Action3}\\&&  \Delta= (M,\alpha), \; \; \; M=1, \cdots, 2n, \; \; \; \alpha = 1, \cdots, m 
\end{eqnarray} 
with,
\begin{equation}
S^{\hbox{{\tiny Max}}}[A^{(\alpha)}_\mu] = -\frac{1}{4} \int d^4 x \delta_{\alpha \beta} f^{(\alpha)}_{\mu \nu} f^{(\beta) \, \mu \nu},
\end{equation}
and where $n$ is now the number of all the vector pairs with the manifestly duality-invariant action, including those that might have been added.

The problem of gauging given rigid symmetries is of course an old question, which has been much studied in the past in the case when the vector fields necessary for achieving the gauging are all described in the free limit by the Maxwell action, i.e., when the starting point is (\ref{Action3}) with only Maxwell terms,
\begin{equation}
S[A^{(\alpha)}_\mu] = S^{\hbox{{\tiny Max}}}[A^{(\alpha)}_\mu] = -\frac{1}{4} \int d^4 x \delta_{\alpha \beta} f^{(\alpha)}_{\mu \nu} f^{(\beta) \, \mu \nu}. \label{SumMaxwells}
\end{equation}
In that case, the consistent gauged formulation is achieved through the Yang-Mills construction \cite{Yang:1954ek}, and the gauge group is a subgroup of the group $SO(m)$ that leaves the second-order action (\ref{SumMaxwells}) invariant.  The uniqueness of the Yang-Mills interactions has been furthermore well established, using a variety of different approaches  \cite{Deser:1963zzc,Berends:1984rq,Wald:1986bj,Barnich:1993pa}. One important result lucidly brought up in \cite{Deser:1963zzc} is that there is no electrically charged massless vector fields. The only way to overcome the inconsistencies encountered in trying to couple charged, i.e., complex, massless vector fields to the electromagnetic field is to go to the Yang-Mills Lagrangian, where all vector fields are on the same footing and where there is no priviledged electromagnetic direction.

But, to our knowledge,  the possibility of gauging electric-magnetic duality,  which would enable the rotation angle $\alpha$ in  (\ref{deltaE0}) and (\ref{deltaB0}) to become a spacetime function instead of a constant, so that being ``electric" or ``magnetic" would be a spacetime-dependent concept, has not been investigated previously.  The possibility of exploring this question becomes available once one uses the first-order action (\ref{Action2}).  If successful, gauging duality would be a novel feature since starting from the standard Maxwell action, one would only gauge a subgroup of $SO(n)$,  not allowing the possibility of gauging transformations of $U(n)$ which are not in $SO(n)$.

We report, however, a negative result.  As we show below, the procedure of modifying the action by replacing the initial abelian curvatures by the non-abelian curvatures stemming out of the would-be new gauge group, does not work because of the impossibility to embed the adjoint action of that group in $U(n)$.

\subsection{Pure first-order action}
To exhibit the obstacle to gauging, we first consider the case where this obstacle appears quite clearly.  This occurs for the pure action (\ref{Action2}) which does not include a standard Maxwell term.

One can view the problem of gauging, and more generally of switching on interactions among the dynamical variables, as a deformation problem, in which one alters both the gauge symmetries and the action keeping the action gauge invariant.  Thus  the number of physical degrees of freedom is not changed by the deformation.  This point of view, systematically developed in \cite{Berends:1984rq,Barnich:1993vg,Barnich:1994db},  is useful in that it clearly organizes the possible obstructions to introducing consistent interactions.  

In our case, the initial gauge symmetry is $[U(1)]^{2n}$ and we want to deform it into a gauge group $G$ of same dimension $2n$.  The group $G$ acts on the potential through the adjoint action.

After the Yang-Mills interactions are switched on, the gauge transformations of the vector fields are deformed to 
\begin{equation}
\delta A^M_\mu = \partial_\mu \Phi^M  + C^M_{\; \; NP} A^N_\mu \Phi^P
\end{equation}
where $ C^M_{\; \; NP}$ are the structure constants of the gauge group $G$ and are defined so as to contain the deformation parameter $g$.  The group $G$ acts also through the adjoint action on the non-abelian curvatures,
\begin{equation}
F^M_{\mu \nu} = f^M_{\mu \nu} + C^M_{\; \; NP} A^N_\mu A^P_\nu
\end{equation} i.e.,
\begin{equation}
\delta F^M_{\mu \nu} = C^M_{\; \; NP} F^N_{\mu \nu} \Phi^P.
\end{equation}

The deformed action is obtained from (\ref{Action2}) by replacing the abelian curvatures by the non-abelian ones 
\begin{equation}
S^{\hbox{{\tiny deformed}}}[A^M_\mu] = \frac{1}{4}  \int dx^0 \, d^3x \left( \sigma_{MN}\epsilon^{ijk} F_{ij}^M \, F_{0k}^N  - \delta_{MN} F_{ij}^M \, F^{N\, ij} \right) \label{Action4}
\end{equation}
It is gauge invariant if and only if the invariance conditions (\ref{InvarianceCond}) on the symplectic form and the metric hold for $$\lambda^M_{\; \; N} =\mu^P \, C^M_{\; \; PN}$$ with $\mu^P$ arbitrary.  That is, if and only if the $2n$ by $2n$ matrices of the adjoint representation of $G$ are matrices of the ($2n$)-dimensional representation of $U(n)$ acting on the $2n$ vector potentials $A^M_\mu$. Explicitly, this imposes
\begin{eqnarray} 
&& \sigma_{QN} C^Q_{\; \; PM}  +\sigma_{MQ}C^Q_{\; \; PN} = 0, \label{InvarianceCondSigma} \\
&&  \delta_{QN} C^Q_{\; \; PM}  +\delta_{MQ} C^Q_{\; \; PN} = 0. \label{InvarianceCondDelta}
\end{eqnarray}

While the second of these invariance conditions poses no problem, the first is the source of the difficulty. Indeed, defining $$D_{MNP} \equiv  \sigma_{MQ} C^Q_{\; \; NP}, $$ one may rewrite (\ref{InvarianceCondSigma}) as
$$D_{NPM} = D_{MPN},$$ i.e., $D_{MNP}$ is symmetric under exchange of its first index $M$ with its last index $P$.  But $D_{MNP}$ is also antisymmetric in its last two indices by definition, $$D_{MNP} = - D_{MPN}$$ which implies $D_{MNP} = 0$ ($D_{MNP} = D_{PNM} = - D_{PMN} = - D_{NMP} = D_{NPM} = D_{MPN} = - D_{MNP} = 0$).  Since $\sigma_{MN}$ is invertible, this yields
$$C^M_{\; \; NP} = 0,$$
which means that $G$ is abelian and that there is no deformation. That is, there is no gauging of any non-abelian subgroup of $U(n)$ and the original abelian gauge symmetry $U(1)^{2n}$ is unchanged. 

A different way to reach the same conclusion is to observe that the $A_0^M$'s drop out from the action (\ref{Action4}) by virtue of the Bianchi identity when the structure constants fulfill the above invariance conditions.  This implies that the $A_0^M$'s should not appear in the field equations.  But one may verify that the Euler-Lagrange derivatives of the kinetic term in (\ref{Action4}) are proportional to $\sigma_{MN}\epsilon^{ijk} D_{0}F_{ij}^M$ and this expression contains the term $\sigma_{MN}\epsilon^{ijk} C^M_{\; \; PQ} A^P_0 F_{ij}^Q$ in which $A^P_0$ explicitly appears.  Hence, since $A^P_0$ must drop out when the invariance conditions are satisfied, this term must be zero and one concludes again that the invariance conditions imply $C^M_{\; \; PQ} = 0$.

\subsection{Mixed Action. Original Noether method}
\label{Noether}

The same conclusion holds for the mixed action (\ref{Action3}) because the rigid symmetry group of linear transformations of the potentials that replaces $U(n)$ is now $U(n) \times SO(m)$.  This group does not mix the $A^M_\mu$'s with the $A^{(\alpha)}_\mu$'s.  Repeating the analysis of the previous subsection, one therefore finds now that the only non-vanishing components of the structure constants are $C^M_{\; \; PQ}$ and $C^\alpha_{\; \; \beta\gamma} $ and further, that the $C^M_{\; \; PQ}$ actually vanish. The only non-trivial Yang-Mills deformations are therefore in the conventional sector, i.e., they are a sum of standard Maxwell actions.

In the original Noether method the obstruction just encountered manifests itself  already at first order in the deformation parameter. To bring this fact to light, consider the gauging of a single $SO(2)$ electric-magnetic duality.  The Noether current that follows from the invariance of the action  (\ref{Action0}) under duality is
\begin{eqnarray}
j^0 &=& \frac{1}{2} \epsilon^{ijk} f^a_{ij} A^b_k \delta_{ab} \\
j^k &=& \epsilon^{ijk} \partial_0 A^a_i \, A^b_j \delta_{ab} - 2 f^{a\, kj} \epsilon_{ab} A^b_j
\end{eqnarray}
This current is coupled to a gauge vector field, which we denote by $B_\mu$, yielding the coupling term,
\begin{equation}g \int d^4x j^\mu B_\mu \label{minimal} \end{equation} where $g$ is the deformation parameter which we do not set equal to unity to keep track of it. The free action for $B_\mu$ is the standard Maxwell action
\begin{equation} S^{\hbox{{\tiny Max}}}[B_\mu] = -\frac{1}{4} \int d^4x G^{\mu \nu} G_{\mu \nu}, \; \; \; \; \; G_{\mu \nu} = \partial_\mu B_\nu -   \partial_\nu B_\mu. \label{MaxForB} \end{equation} We must take (\ref{MaxForB}) for the field $B_\mu$ since we need the component $B_0$ to couple to the component $j^0$ of the current, which is not identically zero.  Thus, the action to zeroth order is  the mixed action (\ref{Action3}) with two fields $A^a_i$ described by the manifestly duality invariant first-order action and one field $B_\mu$ described by the Maxwell action, $M=a= 1,2$, $\alpha = 1$.
 
The minimal coupling term (\ref{minimal}) is invariant under the $U(1)$ gauge symmetry of the field $B_\mu$ up to terms proportional to the field equations for $A^a_\mu$ which can always be compensated by a first-order deformation of the transformations of $A^a_\mu$ under that  symmetry.  
 
 However, the minimal coupling term is not invariant, even on-shell, under the gauge symmetries of $A^a_\mu$.  To cure this problem, one may try to add non-minimal terms involving the curvature $G_{\mu \nu}$.  But although this procedure works when the zeroth order action is a sum of Maxwell actions, yielding the Yang-Mills Lagrangian \cite{Deser:1963zzc}, there is no way here to compensate the non-invariance of (\ref{minimal}).  Indeed, the only candidates that could compensate this non-invariance must be cubic in the fields and involve only one derivative since the minimal term has these properties.  They are therefore, using three-dimensional covariance,  $$\alpha \epsilon^{ijk} \epsilon_{ab} A^a_i A^b_j G_{0k}, \; \; \;  \beta \epsilon_{ab} A^a_i A^b_j G^{ij} $$ but there is no choice of the coefficients $\alpha$ and $\beta$ that make $g \int d^4x j^\mu B_\mu + \alpha \epsilon^{ijk} \epsilon_{ab} A^a_i A^b_j G_{0k} + \beta \epsilon_{ab} A^a_i A^b_j G^{ij}$ invariant on-shell for the gauge symmetries of $A^a_i$.  Hence, the Noether deformation is obstructed already at first-order in $g$.

\subsection{Underlying Group Theory Relationships}

We can reformulate the conditions (\ref{InvarianceCondSigma}) and (\ref{InvarianceCondDelta}) on the structure constants in group theoretical terms as follows. 

The conditions (\ref{InvarianceCondSigma}) and (\ref{InvarianceCondDelta}) respectively express that the adjoint representation of $G$ should be embedded in the symplectic group $Sp(2n, \re)$ and in the orthogonal group $SO(2n)$. Because the adjoint representation is in general not a faithful representation of $G$, this does yield an embedding of $G$ in $Sp(2n, \re)$ and $SO(2n)$, but rather an embedding of $G/Z(G)$ in these groups. Indeed, the kernel of the adjoint representation is equal to the center $Z(G)$, and so the adjoint representation is a faithful representation of the quotient $ G/Z(G)$. 

The impossibility to fulfill the first invariance condition (\ref{InvarianceCondSigma}) is thus the statement that there is no symplectic embedding of the adjoint representation of $G$ unless the adjoint representation and hence $G/Z(G)$ are trivial, that is, unless $G$ is abelian.

\subsection{Relation with pure Yang-Mills}
Our results express that no subgroup of the duality group $U(n)$ can be gauged in the first order, manifestly duality-invariant formulation, not even the subgroups that can be gauged in the second-order formulation.  As we pointed out above in subsection \ref{2.2}, these are subgroups of $SO(n) \subset U(n)$ that act in the same way on $A^{I1}_\mu$ and $A^{I2}_\mu$.

These negative results could seem to lead to a contradiction.  Indeed, one might think that by first gauging the theory in the second order formulation to get the standard pure Yang-Mills action with a subgroup of $SO(n)$ as gauge group,  and then passing to the first-order formulation, one would get a consistent deformation of the first-order action (\ref{Action1}) in which a subgroup of $SO(n)$ is gauged.

It is known, however, that the steps that lead to the manifestly duality-invariant action (\ref{Action1}) in the case of non-interacting $U(1)$-fields cannot be repeated  when the group is non-abelian \cite{Deser:1976iy}. Hence the previous reasoning cannot be applied and no contradiction arises.

\section{Implications for Supergravity}
\label{Supergravity}
\setcounter{equation}{0}

The problem investigated here has some incidence on the gauging of the so-called ``hidden symmetries" of supergravity.  After the discovery that tori dimensional reductions of maximal supergravity exhibits unexpected remarkable symmetries \cite{Cremmer:1978ds,Cremmer:1979up}, the question arose to promote these rigid symmetries into gauge ones.   In four dimensions, the hidden symmetry is $E_{7,7}$ and contains electric-magnetic duality.  This symmetry is thus manifest at the level of the action only if one doubles the vector fields as done above.  

The first gauging of maximal supergravity appeared in \cite{de Wit:1981eq,de Wit:1982ig}.  The general gaugings of that theory were systematically studied more recently in \cite{de Wit:2007mt,deWit:2008ta}.  In all these studies, the starting point is the second order Maxwell action for half of the vector potentials that appear in the discussion.  So, these studies exclude from the very beginning the possibility of gauging the electric-magnetic dualities. Even in the symplectic covariant formalism of \cite{deWit:2005ub}, the constraint that the charges are ``mutually local" is imposed, so that an electric-magnetic duality transformation exists that converts all the charges to electric ones.  In that frame, the gauging is the standard ``electric" gauging of the second order action.

By considering the possibility to gauge the duality transformations that mix ``electric" with ``magnetic", one might have expected to enlarge by a factor of two the available gaugings, since the undeformed gauge group is then $[U(1)]^{2n}$ rather than $[U(1)]^{n}$.  Our negative results, however, show it not to be the case.

\section{Conclusion}
\label{Conclusions}

In this article, we have asked the question: can electric-magnetic duality be gauged? To answer it we have used the standard procedure of (i) Writing the original action in terms of the abelian curvatures; (ii) Replacing the latter by their analog for a non-abelian deformation of the group, and (iii) Demanding that the resulting action be invariant. This procedure is a decantation of the original idea of gauging a global symmetry by iterating the coupling of its Noether current to a compensating field.  We have found that there is no such group deformation. Therefore we conclude that, in the lack of a radically new way to bring in interactions, electric-magnetic duality must be regarded as being only a global symmetry.

\section*{Note added}
After this paper was submitted we have shown in \cite{Bunster:2011aw} that one can extend the proof of the fact that (global) electric-magnetic duality remains a symmetry of the action to the case when one includes couplings to scalar fields defined on the $Sp(2n,\mathbb{R})/U(n)$ coset space.  The duality symetry group is enlarged from $U(n)$ to $Sp(2n,\mathbb{R})$ when the couplings are appropriately chosen \cite{Cremmer:1977tt,Gaillard:1981rj,Breitenlohner:1987dg}.  If one takes the scalar fields on the coset space $G/H$ where $G$ is a subgroup of $Sp(2n,\mathbb{R})$, the duality group is the smaller group $G$. For maximal supergravity in four dimensions,  $G = E_{7,7} \subset Sp(56,\mathbb{R})$ \cite{Cremmer:1978ds,Cremmer:1979up}.   The manifestly $E_{7,7}$ invariant action used in \cite{Hillmann:2009zf,Bossard:2010dq} is, as above,  just the first order form of the action with the Gauss law solved  \cite{Deser:1976iy,Bunster:2011aw},  written in somewhat different notations.

As observed in  \cite{Bunster:2011aw}, gaugings of the Yang-Mills type are not possible either in that case because the kinetic term for the vector fields retains the same form as in (\ref{Action1}) even when the scalar fields are included.

\section*{Acknowledgments} The authors thank the Max-Planck-Institut f\"ur Gravitationsphysik (Albert-Einstein-Institut) for hospitality while this work began to take shape and gratefully acknowledge encouraging comments by Hermann Nicolai.  The Centro de Estudios Cient\'{\i}ficos (CECS) is funded by the Chilean Government through the Centers of Excellence Base Financing Program of Conicyt. M. H.  also  gratefully acknowledges support from the Alexander von Humboldt Foundation through a Humboldt Research Award.  The work of M. H is partially supported by IISN - Belgium (conventions 4.4511.06 and 4.4514.08), by the Belgian Federal Science Policy Office through the Interuniversity Attraction Pole P6/11 and by the ``Communaut\'e Fran\c{c}aise de Belgique" through the ARC program.

\end{document}